# DotMat: Solving Cold-start Problem and Alleviating Sparsity Problem for Recommender Systems

Hao Wang
Ratidar.com
Beijing, China
Haow85@live.com

*Abstract*—Cold-start and sparsity problem are two key intrinsic problems to recommender systems. During the past two decades, researchers and industrial practitioners have spent considerable amount of efforts trying to solve the problems. However, for cold-start problem, most research relies on importing side information to transfer knowledge. A notable exception is ZeroMat, which uses no extra input data. Sparsity is a lesser noticed problem. In this paper, we propose a new algorithm named DotMat that relies on no extra input data, but is capable of solving cold-start and sparsity problems. In experiments, we prove that like ZeroMat, DotMat can achieve competitive results with recommender systems with full data, such as the classic matrix factorization algorithm.

## I. INTRODUCTION

Recommender system is one of the most successful technologies in the big data era. During its early development, internet entrepreneurs believe recommender system would lead to something called Web 3.0, in which net users are recommended information by intelligence. The idea of Web 3.0 reached its climax today when companies such as TikTok, Toutiao.com and Kuai Shou use recommendation as their core product ideas. Recommender systems in the forms of news recommendation, video recommendation, etc. sometimes lead to 30% - 40% increase in web traffic or sales volume.

Recommendation technologies could be dated to the invention of collaborative filtering [1]. From then on, multitudes of research papers have emerged on different research venues. Important inventions such as matrix factorization [2], learning to rank [3][4] and deep learning-based algorithms [5][6][7] have improved the technical accuracy metrics and user experiences step by step. Hardwares supporting recommendation algorithms are becoming more and more complex. Following this trend, most modern recommender systems rely on GPU to run deep-learning programs.

Although recommender systems have evolved for quite a long time, there remain some intrinsic problems to be solved. One notorious problem is the cold-start problem, which deals with the problem of recommending when no information is available for users or items. Nearly all research on the topic utilizes side information to solve the problem. A very famous example is meta learning, which borrows information from other problem domains to tackle the problem.

A notable exception to the research is ZeroMat [8] - an algorithm invented in the year of 2021 that solves the cold-start problem with no input data information. The major supporting foundation of ZeroMat is Zipf Distribution. By assuming user item rating distributed according to Zipf Distribution, researchers were able to construct an algorithm relying on no side information.

One of the approaches of recommender systems combines technologies such as GloVe from NLP with ideas from recommendation to improve recommendation results. Wang [9] modified this idea and created an algorithm named RankMat. The algorithm uses a different probablity distribution family and achieves superior performance in both accuracy metrics and fairness metrics.

In this paper, we propose a new algorithm that borrows the idea from RankMat and ZeroMat to create a more accurate cold-start problem solution. Our method outperforms ZeroMat and heuristic approaches for all parameter values.

## II. RELATED WORK

Recommender systems has been in spotlight in recent years. However, the earliest development of recommender systems could be dated back to decades ago. One of the most successful recommender system algorithms is matrix factorization. There have been a large number of research publications on this topic, and the algorithmic paradigm has been widely adopted in all kinds of internet companies.

The most general framework of matrix factorization is SVDFeature [10], which models the matrix factorization as feature-based dot products of enriched user and item feature vectors. Nearly all matrix factorization variants can be modeled as a particular example of SVDFeature. In 2020, a new algorithm named MatRec [11] was invented to solve the fairness problem of recommender system. MatRec is actually a special case of SVDFeature.

Other matrix factorization variants aiming to solve the fairness problem include Zipf Matrix Factorization [12] and KL-Mat [13]. Both of them rely on regularizing matrix factorization with a fairness penalty term. They are among the multitudes of penalty-based fairness frameworks since 2017.

Alternating Least Squares [14] uses a specific optimization procedure to boost the performance of matrix factorization algorithms. The algorithm have been integrated into open-source machine learning packages such as Spark MLLib.

In 2021, a novel algorithm named RankMat [9] was invented. The algorithm was inspired by GloVeMat [9], a technique carried over into the field from Natural Language Processing. RankMat modifies the underlying probabilistic structure of GloVeMat and achieves superior performance to the classic matrix factorization algorithms.

### III. DOTMAT FRAMEWORK

GloVe is one of the most renown word embedding techniques. The loss function of GloVe is as follows:

$$L = \sum_{i,j=1}^{V} f(X_{ij})(w_i^T \cdot w_j + b_i + b_j - \log X_{ij})^2$$

where f is the weighting function, X is the co-occurrence number of 2 different words, and w and b are parameters to be calculated.

Wang [9] borrowed the idea of GloVe and transformed it into a recommender system named GloVeMat that reformulates the loss function of the algorithm as follows:

$$L = \sum_{i=1}^{m}\sum_{j=1}^{n} (u_i^T \cdot v_j - \log(R_{ij}+1))^2$$

where u and v are user and item feature vectors in matrix factorization paradigm while R is the user item rating value.

It is easy to show that the optimization of L is equivalent to optimize the following loss function:

$$E = e^{w_i^T \cdot w_j + b_i + b_j} - X_{ij}$$

where rank are the popularity ranks of users and items, and they can be approximated by the ranks of users and items in the input data sets. Our new algorithm proposed in this paper is inspired by RankMat. We assume that the user item rating values follow Zipf distribution, namely:

$$R_{i,j} \sim \frac{1}{Rank_{i,j}} \sim U_i^T \cdot V_j$$

Zipf Distribution is a probabilistic distribution introduced first in the area of Linguisitics. The probability depicts the effect that the i-th most popular word in a human language corpus is $\frac{1}{i}$ by occurrence number. We notice Zipf distribution is a fairly accurate depiction of the distribution of the user rating values. This basically means that the user item rating values are distributed proportional to the values themselves, approximated in the form of the dot products of user feature vectors and item feature vectors.

In 2021, Wang [9] modified the framework of GloVeMat and created a new algorithm named RankMat. The formulation of RankMat is as follows :

$$L = ((\frac{1}{\text{rank}_i} \times \frac{1}{\text{rank}_j})^{U_i^T \cdot V_j} - R_{i,j})^2$$

where $rank_i$ and $rank_j$ are the rank of user i and item j respectively.

In DotMat, we modify the formulation of RankMat, and use the following loss function :

$$L = |(\frac{1}{\text{rank}_{i,j}})^{U_i^T \cdot V_j} - \frac{R_{i,j}}{R_{max}}|$$

which, by our discussion on Zipf Distribution, is equivalent to:

$$L = |(U_i^T \cdot V_j)^{U_i^T \cdot v_j} - \frac{R_{i,j}}{R_{max}}|$$

Please notice that unlike RankMat, we use Mean Absolute Error as the evaluation metric. To reconstruct the value of unknown, we have:

$$R_{i,j} = R_{\max} \times (U_i^T \cdot V_j)$$

We use Stochastic Gradient Descent to search for the optimal user feature vector and item feature vector that minimize L:

$$U_i = U_i - \gamma((U_i^T \cdot V_j)^{U_i^T \cdot V_j} \text{sign}((U_i^T \cdot V_j)^{U_i^T \cdot V_j} - \frac{R_{i,j}}{R_{max}})(1 + \log(U_i^T \cdot V_j))V_j)$$

$$V_j = V_j - \gamma((U_i^T \cdot V_j)^{U_i^T \cdot V_j} \text{sign}((U_i^T \cdot V_j)^{U_i^T \cdot V_j} - \frac{R_{i,j}}{R_{max}})(1 + \log(U_i^T \cdot V_j))U_i)$$

We can safely replace with because of Zipf's Law, and we have:

$$U_i = U_i - \gamma((U_i^T \cdot V_j)^{U_i^T \cdot V_j} \text{sign}((U_i^T \cdot V_j)U_i^T \cdot V_j - U_i^T \cdot V_j)(1 + \log(U_i^T \cdot V_j))V_j)$$

$$V_j = V_j - \gamma((U_i^T \cdot V_j)^{U_i^T \cdot V_j} \text{sign}((U_i^T \cdot V_j)_i^T \cdot V_j - U_i^T \cdot V_j)(1 + \log(U_i^T \cdot V_j))U_i)$$

Notice that the computation of user feature and item feature vectors requires no knowledge of user item rating values. Therefore DotMat serves as an ideal choice for cold-start problem if its accuracy and fairness metrics meet the demand.

In addition to the cold-start problem, since DotMat needs no knowledge of historic data, it can be used as a preprocessing step for recommender systems to alleviate sparsity problems. In this way, unknown values of the user item rating matrix can be interpolated from user and item feature vectors before recommendation. We call this hybrid model DotMat Hybrid.

In the Experiment Section, we compare several different recommender system algorithms on open-sourced datasets. The comparison metrics that we use are MAE, i.e., Mean Absolute Error and Degree of Matthew Effect - an idea borrowed from Zipf Matrix Factorization [12].

### IV. EXPERIMENT

We test our algorithm against other algorithms on MovieLens 1 Million Dataset [15] and LDOS-CoMoDa Dataset [16]. MovieLens 1 Million Dataset consists of 6040 users and 3706 movies. LDOS-CoMoDa includes 121 users and 1232 movies. Since comparison results are sensitive to parameter selection, such as the number of user samples and the gradient descent learning step, we do a grid search on gradient descent learning steps, and test 3 different user sample sizes - 100, 1000, and 2000 for MovieLens dataset, and 1 sample size - 100 for LDOS-CoMoDa dataset.

Fig. 1 demonstrates the competitiveness of DotMat when it is used as the preprocessing step before matrix factorization. From the figure, we can also safely draw the conclusion that DotMat is comparable with the classic matrix factorization when gradient learning step is small. Fig. 2 illustrates the comparison of different approaches on fairness metric. It is obvious that DotMat and DotMat Hybrid is competitive with other models.

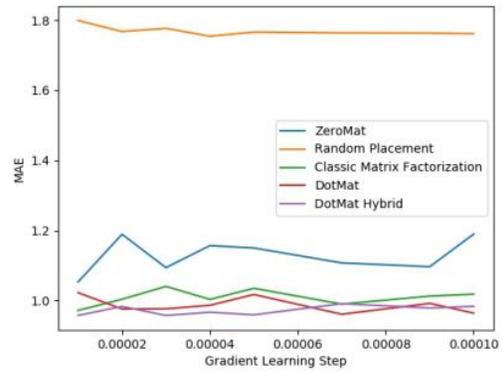

Fig. 3. Comparison of algorithms on MAE when user sample size is 1000.

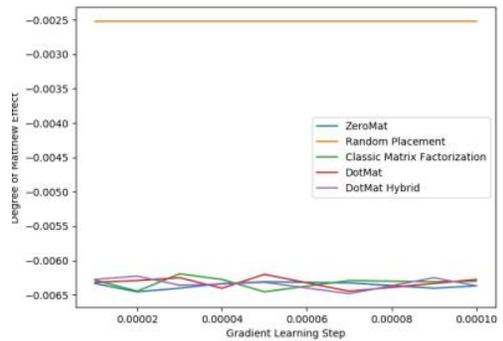

Fig. 4. Comparison of algorithms on fairness metric when user sample size is 1000.

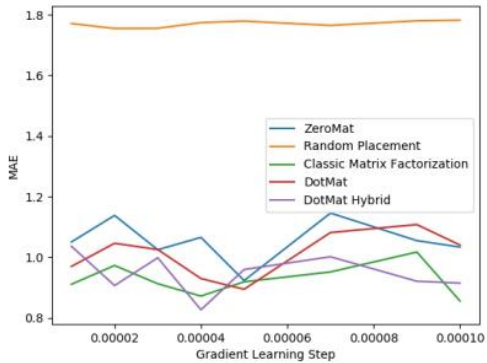

Fig. 1. Comparison of algorithms on MAE when user sample size is 100

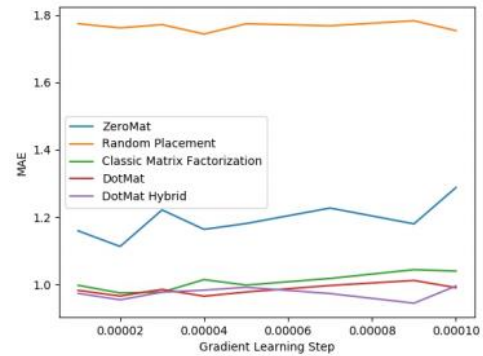

Fig. 5. Comparison of algorithms on MAE when user sample size is 2000

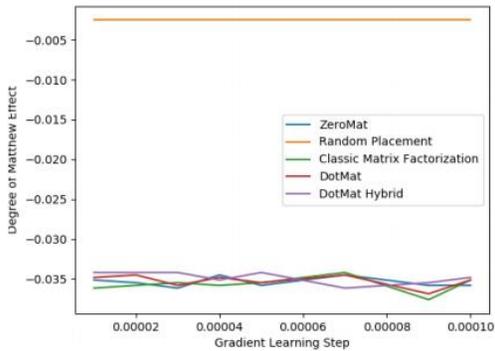

Fig. 2. Comparison of algorithms on fairness metric when user sample size is 100

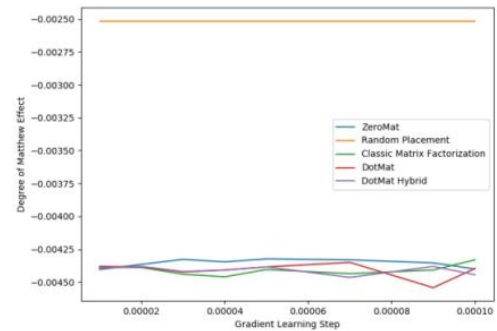

Fig. 6. Comparison of algorithms on fairness metric when user sample size is 2000

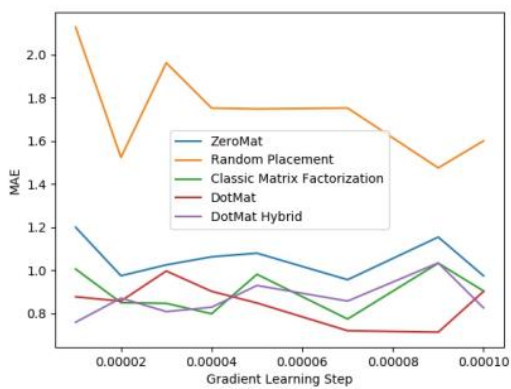

Fig. 7. Comparison of algorithms on MAE when user sample size is 100.

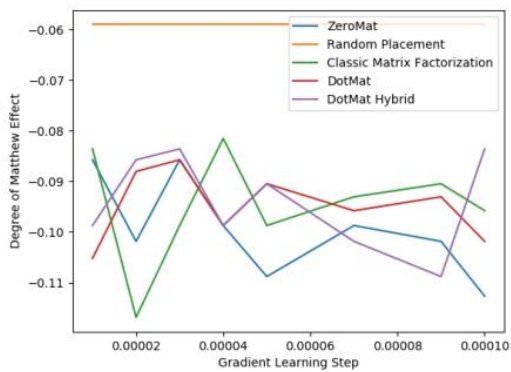

Fig. 8. Comparison of algorithms on fairness metric when user sample size is 100.

From Fig. 3 to Fig. 6, we observe that DotMat as preprocessing step (namely DotMat Hybrid) achieves the best score, while DotMat is competitive with the best performing algorithms.

Fig. 7 and Fig. 8 proves the superiority of DotMat on LDOS-CoMoDa dataset. DotMat and DotMat Hybrid are competitve with ZeroMat on both MAE and fairness metrics.

## V. Discussion

Both ZeroMat and DotMat solve the cold-start problem without any historic input data. However, ZeroMat achieves its best performance (outperforming the classic matrix factorization) when the random user sample size is small while DotMat is competitive for all random sizes. The implications of these 2 algorithms are 2 folds :

1. We are really not doing very well in the recommender system domain, at least when we are using matrix factorization. Because an algorithm without using any data can outperform an algorithm utilizing millions of user item rating values.

2. The human culture evolves into a lock-state as time goes by. Unless exterior interference cuts in, human culture will be predictable at individual level with fairly accurate collective accuracy in a very short historic time.

Since RankMat and DotMat use power law family to model user item rating values and outperform the classic matrix factorization, which is built upon exponential family, we can safely draw the conclusion that power law based loss function is a more accurate framework for matrix factorization.

## VI. Conclusion

In this paper, we propose a new algorithm named DotMat that relies on Zipf Distribution and reformulation of RankMat algorithm. The method does not require any information of user item rating matrix, and is capable of achieving great performance in both accuracy and fairness metrics.

In future work, we would like to explore the reason behind the superiority of DotMat and DotMat Hybrid to classic algorithms. We would like to create even more cold-start and sparisty problem solutions to enhance the performance of recommender systems in the industry.

## Acknowledgment

This paper belongs to a series of independent research work carried out by the author himself. The author would like to thank QingCloud.com for their computing facilities including virtual machines and coupons and compensation plans. The paper was submitted after the author left his previous company, and when he was looking for a job.